\documentclass[twocolumn,pre,showpacs]{revtex4}%
\usepackage{amsmath}%
\usepackage[latin1]{inputenc}%
\usepackage{mathrsfs}

\newcommand{\sech}{\operatorname{sech}}
\newcommand{\Od}[3]{\frac{d^{#1}#2}{d#3^{#1}}}

\newcommand{\PD}[2]{\frac{\partial#1}{\partial#2}}

\newcommand{\dl}{\langle\langle}
\newcommand{\dr}{\rangle\rangle}
\newcommand{\be}{\begin{equation}}
\newcommand{\ee}{\end{equation}}
\newcommand{\ie}{i.~e.\ }
\newcommand{\eg}{e.~g.\ }
\begin{document}
\title{Transverse modulational instability of partially incoherent soliton
stripes}
\author{D.~Anderson, L.~Helczynski-Wolf} \email{lukas@elmagn.chalmers.se}\author{M. Lisak}\author{V.~Semenov}
\affiliation{Department of Electromagnetics, Chalmers University
of Technology, SE-412 96 Göteborg, Sweden} \affiliation{Institute
of Applied Physics RAS, 603950 Nizhny Novgorod, Russia}

\begin{abstract}
Based on the Wigner distribution approach, an analysis of the effect of partial incoherence on the transverse
instability of soliton structures in nonlinear Kerr media is
presented. It is
explicitly shown, that for a Lorentzian incoherence spectrum the
partial incoherence gives rise to a damping which counteracts, and
tends to suppress, the transverse instability growth. However, the
general picture is more complicated and it is shown that the
effect of the partial incoherence depends crucially on the form of
the incoherence spectrum. In fact, for spectra with finite
rms-width, the partial incoherence may even increase both the
growth rate and the range of unstable, transverse wave numbers.
\end{abstract}
\pacs{42.25.Kb, 42.65.Jx,} %Coherence  
\maketitle
\thispagestyle{plain} 
\section{Introduction}
Nonlinear phenomena like self-focusing, collapse,
modulational and transverse instabilities of cylindrical light
beams are some of the most fundamental consequences of the
interplay between linear diffraction and self-phase modulation in
nonlinear Kerr media. Various physical mechanisms, which tend to
suppress such instabilities \eg nonlinear saturation, have been diligently analyzed in a number of works, see references in \cite{YuBook}. These fundamental instability problems have
continued to attract attention in connection with new
scientific and technical developments. There is currently a strong
interest focused on the effects of partial incoherence on
different nonlinear instabilities \cite{Anastassiou,Torres,old,Dragoman,YuBook}. The results
of these studies show that the modulational and collapse
instabilities tend to be suppressed when the waves are partially
incoherent. Recently, the effect of partial incoherence on the
transverse modulational instability of soliton stripes in
nonlinear Kerr media has been investigated, see
\cite{Anastassiou,Torres}. A soliton stripe is a semi-localized
structure, which is of self-trapped soliton form in the $x$
direction, uniform in the $y$ direction and propagates in the $z$
direction. While a 1D soliton is resilient to perturbations, the soliton stripe exhibits instability with respect to
transverse perturbations, \ie perturbations in the $y$ direction,
see \eg \cite{Yuri}. It has been shown, \cite{Anastassiou,Torres}, that when the stripe is
partially incoherent in the $y$ direction, the transverse
modulational instability tends to be suppressed and the break-up
of the stripe, due to the transverse modulational instability, can
be prevented provided the incoherence is sufficiently strong. This behavior is similar to that of the
1D modulational instability. However, analysis of the transverse
modulational instability is more complicated than the
corresponding analysis in the case of 1D modulational instability.
In fact, even in the fully coherent problem, the problem of
finding the growth rate as a function of the wave number of the
perturbations does not have
an explicit analytical solution, cf \cite{Yuri}.\\
In the present work we present an analytical investigation of the
effect of partial incoherence on the transverse instability  of
soliton structures in nonlinear Kerr media. It will be shown
that in the case of a Lorentzian incoherence profile,
the growth rate of the transverse instability can be expressed
simply as the growth rate for the coherent case minus a
stabilizing damping rate due to the partial incoherence. However,
we also show that the case of a Lorentzian profile represents a
very special case and  the effect on the growth rate in a general
case depends crucially on the form of the incoherence spectrum
\cite{prl}. Using a perturbation approach to the dispersion
relation for a general form of the incoherence
spectrum, we show analytically that for weak incoherence spectra
of finite rms-width, the region of instability always widens and the
growth rate is increased in some part of the region. This result
agrees well with a recent numerical study of the transverse
instability of partially incoherent solitons \cite{Torres}, where
the angular spectrum is assumed to have Gaussian form.

\section{The Wigner approach}Our analysis is based on the Wigner approach, which has been shown to be a convenient tool for analyzing the dynamics of partially incoherent light waves, cf \cite{old,Dragoman,YuBook,PRE:Hall}.\\
The starting point of the analysis is the nonlinear Schrödinger
(NLS) equation for the complex wave field, $\psi({\mathbf r},z)$,
describing the two dimensional propagation of a partially coherent
wave in a diffractive  nonlinear Kerr medium, \be
i\PD{\psi}{z}+\frac{1}{2}\nabla^2_\bot\psi+\langle|\psi|^2\rangle\psi=0\label{NLS}
\ee where the bracket $\langle\cdot\rangle$ denotes statistical
average, $z$ is the distance of propagation and ${\mathbf
r}=(x,y)$ denotes the transverse coordinates. The medium response
is here assumed to depend only on the statistically averaged
intensity \ie $I=\langle\psi\psi^*\rangle$. This form of the NLS
equation is valid when the medium response time
is much larger than the characteristic time of the stochastic intensity fluctuations and yet much shorter than the characteristic time of the wave envelope variation.\\
Within the Wigner approach, Eq.(\ref{NLS}) is
transformed into the Wigner-Moyal equation for the corresponding
Wigner function  $\rho({\bf{r},\bf{p}},z)$, viz
 \be \PD{\rho}{z} + {\mathbf
p}\cdot\PD{\rho}{{\mathbf r }} + 2N({\mathbf r },z) \sin
\left(\frac{1}{2} \frac{\overleftarrow{\partial}}{\partial
{\mathbf r }}\cdot \frac{\overrightarrow{\partial}}{\partial
{\mathbf p }} \right)\rho({\mathbf r},{\mathbf p },z)=0. \ee The
Wigner distribution is determined by the stochastic properties of
$\psi({\bf r},z)$ and conversely
$N=\langle|\psi|^2\rangle=\int\rho d{\mathbf p}$ is the average
field intensity. In the present application we consider a
background solution in the form of a soliton stripe, \ie a
semi-localized structure, which constitutes a self-trapped soliton
form in the $x$ direction, is uniform in the $y$ direction and
propagates in the $z$ direction. This structure is assumed
partially incoherent in the $y$ direction. The corresponding
intensity and the concomitant Wigner distribution are
\be
N_0(x)=\sech^2(x)\ee and \be
\rho_0(x,{\mathbf p})=\frac{2\sin(2x p_x)}{\sinh(2x)\sinh(\pi
p_x)}G(p_y)\equiv R_0(x,p_x)G(p_y), \ee respectively, where
$G(p_y)$ characterizes the spectrum of the partial incoherence in
the transverse direction. In order to analyze the stability of
this background solution, we consider the dynamics of a small
perturbation by writing $\rho=\rho_0(x,{\mathbf
p})+\rho_1({\mathbf r},{\mathbf p},z)$, where $\rho_1\ll\rho_0$.
The linear evolution of the small perturbation $\rho_1$ is then
governed by the equation
\begin{align}
&\PD{\rho_1}{z} + {\mathbf p}\cdot\PD{\rho_1}{{\mathbf r }} + 2
N_0 \sin \left(\frac{1}{2}
\frac{\overleftarrow{\partial}}{\partial x}
\frac{\overrightarrow{\partial}}{\partial p_x}
\right)\rho_1({\mathbf r },{\mathbf p },z) + \nonumber \\&
2n_1({\mathbf r },z)\sin\left(\frac{1}{2}
\frac{\overleftarrow{\partial}}{\partial {\mathbf r }} \cdot
\frac{\overrightarrow{\partial}}{\partial {\mathbf p }}
\right)\rho_0(x,{\mathbf p}) =0. \label{main}
\end{align}
where $n_1=\int \rho_1 d{\mathbf p}$. When considering the
transverse modulational instability, the perturbations can be
assumed to be  described by harmonic variations, \ie $n_1({\mathbf
r},z)=n(x)\cos(ky)\exp(\Gamma z)$, where $k$ is the wave number of
the transverse perturbation. With this ansatz for the
perturbation, Eq.(\ref{main}) can be rewritten in the compact form
\be \frac{\partial\rho_1}{\partial z} + {\mathbf
p}\cdot\PD{\rho_1}{{\mathbf r }} + 2N_0\hat{S}\rho_1 +
G_+\;n\hat{S}R_0 + \frac{G_-}{k}\PD{n_1}{y}\hat{C}R_0=0
\label{compactform}\ee where we have introduced the operators
\begin{equation*}
\hat{S}=\sin\left(\frac{1}{2}
\frac{\overleftarrow{\partial}}{\partial x}
\frac{\overrightarrow{\partial}}{\partial p_x}
\right),\hspace{1cm} \hat{C}=\cos\left(\frac{1}{2}
\frac{\overleftarrow{\partial}}{\partial x}
\frac{\overrightarrow{\partial}}{\partial p_x} \right)
\end{equation*}
and used the notations $G_+=G(p_y+k/2) + G(p_y-k/2)$ and
$G_-=G(p_y+k/2) - G(p_y-k/2)$.\\
The solution of Eq.(\ref{compactform}) can be represented as
\be\rho_1 =[U(x,{\mathbf p })\cos(ky) + V(x,{\mathbf p })\sin(ky)]\exp(\Gamma
z),\ee where the unknown functions U and V satisfy the equations
\be\begin{aligned}
&\Gamma U + p_x \PD{U}{x} + 2N_0\hat{S}U+kp_yV = -G_+\;n\hat{S}R_0,  \\
&\Gamma V + p_x \PD{V}{x} + 2N_0\hat{S}V-kp_yU =
G_-\;n\hat{C}R_0.
\end{aligned}\label{gammaUV}\ee
This equation system has to be solved subject to the consistency
conditions $\int U\,dp_x\,dp_y=n(x)$ and $\int V\,dp_x\,dp_y=0$.

\section{The case of Lorentzian incoherence spectrum}
For the  development of our analysis it is useful to first
reconsider the case of a fully coherent wave. The transverse
coherence spectrum is then a Dirac delta function \ie
$G(p_y)=\delta(p_y)$. The earlier introduced notations $G_+$ and
$G_-$ now become a sum and a difference, respectively, of two
translated delta functions. The $p_y$ dependence of the U and V
functions can be expressed in similar manner, \ie $
U=[\delta(p_y+k/2)+\delta(p_y-k/2)]\tilde{u}({\mathbf r },p_x)$
and $V=[\delta(p_y+k/2)-\delta(p_y-k/2)]\tilde{v}({\mathbf r
},p_x)$. The combinations of delta functions  now appearing in
Eq.(\ref{gammaUV}) can be shown to be separable, and the resulting
system of equations reduces to
\be\begin{aligned}
&\hat{L} \tilde{u} -\frac{k^2}{2}\tilde{v} = -n\hat{S}R_0 \\
&\hat{L} \tilde{v} +\frac{k^2}{2}\tilde{u} = n\hat{C}R_0
\end{aligned}\label{tildeUtildeV}\ee
where we have introduced a $p_y$-independent operator $\hat{L}$
defined as $\hat{L}=\Gamma+p_x\partial/\partial x+2N_0\hat{S}$.
Eqs.(\ref{tildeUtildeV}) can be combined into a single equation for
$\tilde{u}$, which reads \be \tilde{u}
-k^2\hat{P}^{-1}\{\hat{L}\}n\hat{C}R_0
+2\hat{L}\hat{P}^{-1}\{\hat{L}\}
n\hat{S}R_0=0.\label{coherentfall}\ee where $\hat{P}^{-1}$ denotes
the inverse of the operator
$\hat{P}\{\hat{L}\}=k^4/4+(\hat{L})^2$ and
curly brackets denote the argument of the operator. The solution of
the eigenvalue problem ($\Gamma=\Gamma(0,k)$) cannot be found analytically, and resort
must be taken to approximate analytical techniques and/or
numerical computations, cf \cite{Rubenchik, Anderson}. As an example,
a derivation inspired by direct variational methods is given in the Appendix.\\
With this result  in mind for later comparison, we turn back to
the partially incoherent problem. In the same way as for the
coherent case, we can eliminate the function $V$ in
Eq.(\ref{gammaUV}) to obtain
\be \hat{L}^2U +
kp_y(kp_yU + G_-\;n\hat{C}R_0)=
-\hat{L}G_+\,n\hat{S}R_0. \ee
Integrating this equation over $p_y$-space we obtain:
\be\begin{split} &\int_{-\infty}^\infty U\,dp_y
+\int_{-\infty}^\infty\hat{A}^{-1}G_-\; kp_yn\hat{C}R_0 \,dp_y
\\&+ \int_{-\infty}^\infty\hat{A}^{-1}G_+ \hat{L}n\hat{S}R_0
,dp_y=0,
\end{split}\label{integration}\ee
where yet another new operator, $\hat{A}$, has been introduced.
It is defined as $\hat{A}=[k^2p_y^2 +\hat{L}^2]$ and $\hat{A}^{-1}$
denotes its inverse. Since the p$_y$ dependence in the operator
$\hat{A}^{-1}$ is multiplicative, some important simplifications
can be made. For instance, in the second integral of
Eq.(\ref{integration}), the ordering of the terms may be
interchanged as $\hat{A}^{-1}G_-\; p_y n\hat{C}R_0=G_-\;
p_y\hat{A}^{-1} n\hat{C}R_0$. For the
subsequent analysis we need the eigenvalue rather than the
operator itself since $\hat{A}^{-1}n\hat{C}R_0=\sum_m
a_m^{-1}(p_y)c_mn \hat{C}R_0$. The eigenvalue $a_m^{-1}$ corresponding to the
operator $\hat{A}^{-1}$ is given by
$a_m^{-1}=1/(\lambda_m^2+k^2p_y^2)$
where, in turn, $\lambda_m^2$ is the eigenvalue of the $\hat{L}^2$ operator.\\
We will now assume the incoherence spectrum to have a Lorentzian
profile, $G(p_y)=p_0/[\pi(p_y^2+p_0^2)]$, with the characteristic width $p_0$. This assumption has the
important consequence that the integrals appearing in
Eq.(\ref{integration}) can be evaluated explicitly to yield:
\begin{align}
\int_{-\infty}^\infty a_m^{-1}G_- p_y \,dp_y &= \frac{-1}{k\left[\left(k/2\right)^2+\left(p_0+\lambda_m/k\right)^2\right]}\\
\int_{-\infty}^\infty a_m^{-1}G_+ \,dp_y
&=\frac{2(\lambda_m+kp_0)}{k^2\lambda_m\left[\left(k/2\right)^2+
\left(p_0+\lambda_m/k\right)^2\right]}.
\end{align}
Thus the dispersion equation, Eq.(\ref{integration}), can be
expressed in the following form
\begin{align} \int_{-\infty}^\infty U\,dp_y -& k^2
\hat{P}^{-1}\{kp_0+\hat{L}\}n\hat{C}R_0 \nonumber\\ %(kp_0+\hat{L})
&+2\left(kp_0+\hat{L}\right)\hat{P}^{-1}\{kp_0+\hat{L}\} %(kp_0+\hat{L})
n\hat{S}R_0=0. \label{finalDisp}\end{align}
A comparison of the two dispersion relations (the one for the
coherent, Eq.(\ref{coherentfall}), and the one for the partially
incoherent, Eq.(\ref{finalDisp}), case) shows that the only
difference between the two is   the shift in the argument of the
$\hat{P}$ operator; the argument $\hat{L}$ is replaced by
($\hat{L} + kp_0$) in the partially incoherent case. Equivalently, since
$\hat{L}=\Gamma+p_x\partial/\partial x+2N_0\hat{S},$ this implies
that $\Gamma(0,k)=\Gamma(p_0,k)+kp_0$, where $\Gamma(p_0,k)$
denotes the growth rate of the partially incoherent case. Thus we
finally come to the important conclusion that, for Lorentzian
shaped incoherence spectrum, the role of the partial incoherence
on the transverse modulational instability of a soliton stripe can
be expressed in exactly the same form as for the 1D modulational
case \cite{PRE:Hall}, viz simply as a stabilizing damping
according to \be\Gamma(p_0,k)=\Gamma(0,k)-kp_0,\ee where
$\Gamma(0,k)\approx k\sqrt{3-k^2}/2$, cf Appendix. This implies
two things: the instability is suppressed by the incoherence for
all wave numbers in the range $[0,k_c]$, where the cut-off wave
number, $k_c$, is given by $k_c=\sqrt{3-4p_0^2}$ and secondly the
range of instability decreases monotonously with increasing
incoherence. However, this simple monotonously suppressing effect
of the partial incoherence on the transverse modulational
instability is not of a general nature. An indication of this was
found in \cite{Torres}, where numerical investigations were made
using Gaussian as well as Lorentzian coherence spectra. Somewhat
counter-intuitively it was found that for the case of a Gaussian
spectrum,  increasing  incoherence  actually increased the range
of modulationally unstable wave numbers and increased the growth
rate in part of the unstable region. Only for sufficiently strong
incoherence did the unstable wavelength range start to shrink and
the growth rate to decrease and to ultimately vanish. Thus, it
seems that the properties of the transverse modulational
instability depend crucially on the form of the incoherence
spectrum. That this indeed is so will be shown analytically in the
subsequent paragraph.

\section{Results for a general incoherence spectrum}
In general, a complete analytical solution of
Eqs.(\ref{gammaUV}) seems impossible to find. However, important
information about the properties of the solution can be obtained
by considering certain moments of the equations. For this purpose,
we integrate the coupled equations for $U$ and $V$ over $x$ and
$p_x$. This yields
\be\begin{aligned}
&\Gamma\dl U\dr+kp_y\dl V\dr=0\\
&\Gamma\dl V\dr-kp_y\dl U\dr=G_-\dl nR_0\dr,
\end{aligned}\label{UV}\ee where double brackets
$\dl\cdot\dr$ denote integration over $x$ and $p_x$. The
consistency condition for the real part of the perturbation can
conveniently be expressed as \be \int_{-\infty}^\infty\dl U\dr
dp_y=\dl n\delta(p_x)\dr.\label{consistent}\ee Thus, solving for
$\dl U\dr$ from Eq.(\ref{UV}) and inserting this into
Eq.(\ref{consistent}), we obtain the dispersion relation for the
transverse instability of incoherent solitons in the form
\be\int_{-\infty}^\infty\frac{kp_yG_-}{\Gamma^2+(kp_y)^2}dp_y
=-\frac{1}{Q}=-\frac{\dl n\delta(p_x)\dr}{\dl
nR_0\dr}\label{Disp}.\ee We underline that $G_-$ is determined by
the coherence properties of the soliton background solution, but
that the parameter $Q$ may
depend on the coherence spectrum. Nevertheless, the result
expressed by Eq.(\ref{Disp}) is completely general and is valid
for arbitrary form of the coherence spectrum. We emphasize that
Eq.(\ref{Disp}) is of the same form as the dispersion relation for
the modulational instability of a partially coherent, but
homogeneous, background, cf \cite{PRE:Hall,prl}, in which case the
parameter $Q$ is easily determined to be $Q=1$. On the other hand,
for the transverse instability of a partially incoherent soliton
stripe, the proper value of $Q$ can not be easily found, although we may state that $Q<1$. For the
special case of a Lorentzian spectrum studied above, we can take
one step further in Eq.(\ref{Disp}) to obtain a dispersion
relation \be (\Gamma+kp_0)^2=(2Q-k^2/2)k^2/2\label{Gamma} \ee
where, however, the Q factor  still remains to be determined. The analysis
of the previous section and the result of the Appendix indicate that,
 $Q\approx 3/4$, independently of the degree of
incoherence i.e. independently of
$p_0$.\\
In order to pursue this line of analysis for general forms of
incoherence spectra, we will assume weak partial incoherence in
the sense that the incoherence spectrum is very narrow \ie $p_0\ll k$. The
integral of Eq.(\ref{Disp}) may then be evaluated approximately
for any (well behaved) incoherence spectrum $G(p_y)$. This implies
that the function $F(p_y)=k p_y/(\Gamma^2+k^2p_y^2)$ multiplying
$G_-$ in the integral can be expanded around the shifted wave
numbers $\pm k/2$ to yield
\begin{align} \int_{-\infty}^\infty
&F(p_y)[G(p_y+k/2)-G(p_y-k/2)]dp_y\approx\nonumber\\
&-\frac{k^2}{\Phi}+\frac{k^4}{\Phi^2}\left(3-\frac{k^4}{\Phi}\right)p_{\text{rms}}^2,\label{approximate}
\end{align}  where $\Phi=\Gamma^2+k^4/4$ and we have defined
$p_{\text{rms}}^2\equiv\langle x^2\rangle$ as the rms-width of the spectrum
\be \langle x^2\rangle=\int_{-\infty}^\infty x^2
G(x)dx\left/\rule{0pt}{.5cm}\right. \int_{-\infty}^\infty G(x)dx.
\ee
The dispersion relation given by Eq.(\ref{Disp}) then becomes
\be
\Gamma^2=k^2\left(Q-\frac{k^2}{4}\right) -\frac{k^4
Q}{\Phi}\left(3-\frac{k^4}{\Phi}\right)p_{\text{rms}}^2 \ee
Since the incoherence is assumed weak, we will assume that the dispersion relation given by Eq.(\ref{approximate}) may
be simplified perturbatively by taking $Q$ equal to its coherent
value $Q_c$ and replacing $\Phi=\Gamma^2+k^4/4\approx k^2Q_c$ in the incoherently induced correction term. This yields
\be \Gamma^2\approx k^2\left[
\left(Q_c-\frac{k^2}{4}\right)-\left(3-\frac{k^2}{Q_c}\right)p_{\text{rms}}^2\right]
\ee
From this approximate expression for the growth rate, we can
draw two important conclusions, valid for arbitrary (but narrow)
incoherence spectra with finite rms-width: (i) the instability
tends to be suppressed for all wave numbers in the range
$0<k^2\leq 3Q_c$, whereas in the region $3Q_c<k^2<4Q_c$, the growth
rate is enhanced by the partial incoherence. (ii) the critical
(non-zero) wave number, $k_c$, at which the growth rate goes to
zero, increases and is given by $k_c^2\approx 4(Q_c+p_{\text{rms}}^2)\approx 3+4p_{\text{rms}}^2$.
These analytical
results agree well with what was  obtained by Torres et al. \cite{Torres} using numerical computations.\\
On the other hand, these results are in contradiction
with the results obtained in the previous section for the case of a Lorentzian
spectrum. There it was found that (i) the growth rate decreased
for all wave numbers, (ii) the cut off wave number, $k_c$,
monotonously decreased with increasing incoherence. The
explanation of this apparent contradiction is that the analysis of
this section excludes spectra, which, like the Lorentzian, do
not have a finite rms-width. A direct implication of this result is
that the effect of partial incoherence depends crucially on the
form of the incoherence spectrum, even to the extent that in some
wavelength range the instability may even be enhanced by the
incoherence. As demonstrated in \cite{Torres}, for increasing
incoherence, the range of unstable wave numbers first increases,
but then eventually shrinks until finally the instability is
completely quenched. This complete behavior is outside the range of
validity of the perturbation analysis presented in the current
section.

\section{Conclusion}
The present analysis has, in some detail, considered the effect of
partial incoherence on the transverse modulational instability of
soliton stripes. We have shown that, for a Lorentzian form of the
incoherence spectrum, the effect of partial incoherence on the
transverse instability agrees qualitatively with the corresponding
result derived for the case of 1D modulational instability; the growth rate decreases monotonously for increasing
partial incoherence. However, the Lorentzian form is a very
special case in the sense that although it has the nice property
of being analytically integrable, it does not have a finite
rms-width. Our analysis of general spectra with finite rms-widths
shows quite a different qualitative behavior of the growth rate
for weak increasing incoherence. The growth rate is found to
decrease for transverse wave numbers in the range $0<k<k_*$, but
to increase in the complementary range $k_*<k<k_c$, where $k_c$ is
the cut off wave number of the instability and $k_*$ is a
characteristic transition wave number. In addition, it is found
that $k_c$ does in fact increase.  These analytical results agree well with
numerical simulations performed by \cite{Torres} as well as with
previous analytical work of ours for the simpler case of the 1D
modulational instability, \cite{prl}.

\appendix*
\section{}
The dispersion relation for the transverse modulational
instability  cannot be determined analytically even in the
coherent case and several different approximations have been
presented, cf \cite{Rubenchik,Laedke,Anderson}. 
We will here give a simple, accurate and as far as we know new,
approximation using a direct variational approach. Linearization
of the 2D coherent NLS equation, given in Eq.(\ref{NLS}), around the stationary solution
$\psi=\sech x\exp(iz/2)$ gives rise to two coupled equations for
the real $u(x)$, and imaginary $v(x)$, parts of the perturbed wave
field. Inserting the assumed variations in $y$ and $z$ for the
modulational perturbations (\ie $u,v\propto \exp(iky+\Gamma z)$),
these equations become: \be\Gamma u=\hat{L}_1v\quad;\quad \Gamma
v=-\hat{L}_2u \label{classical}\ee where the operators $\hat{L}_1$
and $\hat{L}_2$ are self-adjoint and defined by 
\be\begin{aligned}
&\hat{L}_1=-\frac{1}{2}\Od{2}{}{x}+\frac{1}{2}(1+k^2)-\sech^2x, \\
&\hat{L}_2=\hat{L}_1 -2\sech^2x. 
\label{A2}\end{aligned}\ee
Eqs.(\ref{classical}) can be reformulated as a variational problem
corresponding to the Lagrangian $\mathscr{L}=\frac{1}{2}v\hat{L}_1v -\frac{1}{2}u\hat{L}_2u-\Gamma uv$. An ansatz is made for the functions $u$ and $v$ as $u=\alpha
\phi_2$, $v=\beta\phi_1$, where $\phi_1$ and $\phi_2$  are  trial
functions and  $\alpha$ and $\beta$ are the variational
parameters. Inserting this ansatz into the variational integral,
we find
$\langle\mathscr{L}\rangle=\frac{1}{2}\alpha^2\langle\phi_1|\hat{L}_1|\phi_1\rangle-
\alpha\beta\Gamma\langle\phi_1|\phi_2\rangle- \frac{1}{2}\beta^2\langle\phi_2|\hat{L}_2|\phi_2\rangle$, where brackets $\langle\cdot\rangle$ denote integration over $x$.
The variational equations with respect to $\alpha$ and $\beta$
give rise to a linear system of equations for these parameters. A
nontrivial solution of the system requires its determinant to
vanish, giving the following dispersion relation \be
\Gamma^2=-\frac{\langle\phi_2|\hat{L}_2|\phi_2\rangle
\langle\phi_1|\hat{L}_1|\phi_1\rangle}{\langle\phi_1|\phi_2\rangle^2}.\ee
With the intuitive choice of the trial functions as equal to the
eigenfunctions of the operators $\hat{L}_1$ and $\hat{L}_2$, \ie $\phi_1=\sech x$ and
$\phi_2=\sech^2 x$ respectively, the dispersion relation for the coherent case
of the transvere instability becomes \be
\Gamma^2(0,k)\approx\Gamma^2=k^2(3-k^2)\frac{8}{3\pi^2}\approx\frac{k^2}{4}(3-k^2).\ee

\end{document}